\documentclass[twocolumn,showpacs,amssymb,aps, prl]{revtex4}
\usepackage{amsmath}
\usepackage{graphicx}
\usepackage{amssymb}
\usepackage{mathrsfs}
\newtheorem{thm}{Theorem}
\def\adm{{\mathbf m}_{\textrm{ADM}}}
\def\by{{\mathbf m}_{\textrm{BY}}}
\def\ly{{\mathbf m}_{\textrm{LY}}}
\def\hawking{{\mathbf m}_{\textrm{H}}}

\def\W{\mathbf{W}}
\newcommand{\lie}{\mathcal{L}}
\newcommand{\ud}{\mathrm{d}}

\newcommand{\beq}{\begin{eqnarray}}  \newcommand{\eeq}{\end{eqnarray}}

\begin{document}

\title{Localized Penrose inequality for the Liu-Yau mass in spherical symmetry}

\author{Fei-hung Ho}\email{fei@ntnu.edu.tw}
\affiliation{
Department of Mathematics and Science, College of International Studies and Education for Overseas Chinese Students,
National Taiwan Normal University, Taipei, Taiwan
}
\author{Jian-liang Liu}\email{tendauliang@gmail.com}\affiliation{Shanghai Center for Mathematical Sciences, Fudan University, Shanghai, China}

\author{Naqing Xie}\email{nqxie@fudan.edu.cn}
  \affiliation{School of Mathematical Sciences, Fudan University, Shanghai, China}

\date{\today}

\begin{abstract}
For an asymptotically flat initial data, the Penrose inequality gives a lower bound of the Arnowitt-Deser-Misner total mass of a spacetime in terms of the area of certain surfaces representing black holes. This is a deep and beautiful refinement of the famous positive mass theorem and it plays an important role in the study of gravitational collapse. Gravitational collapse can also happen if sufficient mass is concentrated into a finite region. This motivates us to seek a localized version of the Penrose inequality. In this Letter, we successfully make a precise statement of this form for the Liu-Yau quasi-local mass in spherical symmetry.
\end{abstract}

\pacs{04.20.Cv, 04.20.Dw}
\keywords{}
\maketitle

{\it\noindent Introduction.} The Penrose inequality \cite{Pen1}, a generalization of the positive mass theorem \cite{S+YI,S+YII,Wi}, gives a lower bound of the total mass of a spacetime in terms of the area of certain surface that represents black hole. The original physical arguments led Penrose to conjecture such an inequality are usually referred to as the Cosmic censorship hypothesis \cite{Pen3}. On a heuristic level, if the inequality failed, that would indicate likely failure of the Cosmic censorship. Therefore, a mathematically rigorous proof, or a physically falsifiable disproof, of the Penrose inequality is of vital importance in general relativity.

The inequality for a time symmetric initial data is an important special case which is called the Riemannian Penrose inequality. Precisely, let $(M^3,g)$ be a complete, asymptotically flat Riemannian manifold which has nonnegative scalar curvature. Assume that the boundary $\Sigma=\partial M$ is the outermost apparent horizon. The Riemannian Penrose inequality states that
\beq\label{RPI}
\adm \geq \sqrt{|\Sigma|/16\pi}.
\eeq
Here $\adm$ is the Arnowitt-Deser-Misner total mass \cite{ADM} of the initial data $(M^3,g)$, and $|\Sigma|$ is the area of the outermost apparent horizon. Usually it is also accompanied by a rigidity statement saying that the equality holds if and only if $(M^3,g)$ is isometric to a spatial Schwarzschild space outside its horizon. This inequality was proved by Huisken and Ilmanen for one black hole on the machinery of the weakly defined inverse mean curvature flow \cite{HI}, and independently by Bray for multiple black holes using conformal flow \cite{Br}. The proof of the Riemannian Penrose inequality is very difficult but a bit of a mystery. It involves many techniques developed in geometric analysis and partial differential equations. Recently, Ohashi, Shiromizu, and Yamada reinterpreted and modified Bray's conformal flow and their argument turned out to reveal the physical insight between the metric flow and the gravitational collapse \cite{OSY}. In particular, they showed that the normalized conformal flow corresponds to a virtual time evolution of gravitational collapse, satisfying a non-vacuum Einstein equation.

For a general non-time symmetric initial data, with the analysis of the Jang equation \cite{J}, Schoen and Yau successfully proved the spacetime version of the positive mass theorem under the dominant energy condition \cite{S+YII}. There exist various serious attempts to the formulation of the spacetime Penrose inequality. We refer to \cite{Mar} for surveys in this topic. For instance, Bray and Khuri recently proposed a version of the Penrose inequality in terms of the generalized trapped surfaces instead of the (weakly) outer trapped surfaces \cite{BK}. They showed that if the generalized Jang equation system
admits solutions with appropriate boundary behavior, then such a Penrose inequality holds. Unfortunately, Carrasco and Mars discovered a counterexample on axially symmetric slices in the Kruskal spacetime which violates this inequality \cite{CM}. Instead, Bray and Khuri then conjectured the Penrose inequality for the time-independent apparent horizons \cite{BK}. The full spacetime version of Penrose inequality still remains open and is in the running.

While in spherical symmetry, Malec and \'{O} Murchadha successfully established such a Penrose inequality \cite{MO}. They proved:

\noindent {\bf Theorem (Malec-\'{O} Murchadha, \cite{MO})} {\it Let $\Sigma_t$ be a partial maximal Cauchy hypersurface that extends outwards from an outermost future
(past) apparent horizon situated at a sphere $S$. Assume the dominant energy condition in $\Sigma_t$, and let the ADM
mass be $m$. Then the areal radius $R =\sqrt{|S|/4\pi}$ of the future
(past) apparent horizon must be less than the
Schwarzschild radius $2m$.}

It should be pointed out that the same result holds true also for any non-maximal
Cauchy hypersurface, under the remaining conditions as above. (cf. \cite[Page 6933, Remark under Theorem]{MO})

It is widely believed that if sufficient mass is concentrated into a finite region, gravitational collapse will happen. One would seek a localized statement of the Penrose inequality (\ref{RPI}). It still leaves the ambiguity of the ``mass''. At any point, we can always choose the coordinates so that the metric is Minkowski with all first order derivatives vanishing. This makes the concept of the local
energy density ill-defined. We feel it natural to use one of the quasi-local masses. It captures the idea to quantify the total mass (both gravitational and matter) inside a closed 2-surface. There are already many candidates in the literature. For instance, we have the Brown-York mass \cite{BY}, the Bartnik mass \cite{Ba}, the Misner-Sharp mass \cite{MS}, the Hawking mass \cite{Ha}, the Kijwoski mass \cite{Kij}, the Chen-Nester-Tung mass \cite{CNT}, the Liu-Yau mass \cite{LY}, and the very recent Wang-Yau mass \cite{WY} and Zhang mass \cite{Zh}. A comprehensive survey
is given by Szabados in \cite{Sz}. And we believe that each existing candidate may be a mixture of ``good'' and ``bad'' characteristics and the hunting season for an optimal quasi-local mass is still open.

Happily, in spherically symmetric geometries, we are able to show the precious value of the Liu-Yau quasi-local mass by providing a localized Penrose inequality.

We have a spherically symmetric spacetime. Let $\Omega$ be a spherical, compact, oriented, three dimensional spacelike time slice with boundary $\Sigma$. Suppose that the boundary $\Sigma$ is the disjoint union of two pieces, $\Sigma=\Sigma_{\textrm{O}} \cup \Sigma_{\textrm{H}}$ satisfying the following conditions
(i) The outer boundary $\Sigma_{\textrm{O}}$ is a round 2-sphere; (ii) The inner boundary $\Sigma_{\textrm{H}}$ is an outermost future apparent horizon, which is also a round sphere.
We show that

\begin{thm}[Localized Penrose Inequality]
Assume further that the dominant energy condition holds on $\Omega$. Then the areal radius of the outermost future apparent horizon $\Sigma_{\textrm{H}}$  must be less than twice of the Liu-Yau quasi-local mass of the outer boundary $\Sigma_{\emph{\textrm{O}}}$.
\end{thm}

It is equivalent to reformulate the above theorem as
\begin{equation}\label{lpi}\ly(\Sigma_{\textrm{O}}) \geq \sqrt{|\Sigma_{\textrm{H}}|/16\pi}.
\end{equation}There is a fundamental requirement for a quasi-local mass that marginally trapped surfaces with $S^2$ topology ``enclose'' the irreducible mass $\sqrt{|S|/16\pi}$. This novel inequality positively supports the Liu-Yau mass in this aspect.

{\it\noindent Analysis and Proof.} A spherical spacetime can be described as
\begin{equation*}
\ud s^2=-N^2(t,r)\ud t^2+ \lie^2(t,r)\ud r^2 + R^2(t,r) \Big(\ud \theta^2+\sin^2\theta\ud \varphi^2 \Big).\end{equation*}

We shall follow the notations in \cite{GO}. Without loss of generality, we can take the proper time parameter so that $N(t,r)\equiv 1$. The spatial geometry at constant time slice has two radial functions $\lie(r)$ and $R(r)$. We denote by prime the derivative with respect to the proper radius defined by $\ud l=\lie \ud r$ and denote by dot the derivative with respect to the proper time. The advantage that radial derivatives are taken with respect to $l$ is that $\lie$ no longer appears in the constraint equations.

For a spacelike closed 2-surface $\Sigma$ in the time slice, we denote by $\vec{v}$, the outward spacelike unit normal, and denote by $\vec{u}$ the future directed timelike unit normal of the time slice in the spacetime. Associated with each of them is an expansion. The expansion along $\vec{v}$, called $k$, is the mean curvature of $\Sigma$ embedded in the time slice. The other one along the $\vec{u}$ direction, called $p$, is the 2-trace of the extrinsic curvature $K_{ab}$ of the time slice in spacetime. Equivalently, if we write in terms of null expansions with respect to the outgoing rays, that is the fractional rate of change of the 2-area when dragged along the two null normals, one has $\theta_+=k+p$ and $\theta_-=k-p$. We further assume that the mean curvature vector $\vec{H}$ of $\Sigma$ is spacelike so that the associated Liu-Yau mass is well-defined, i.e. $k^2-p^2>0$.

In spherical symmetry, the extrinsic curvature $K_{ab}$ of the time slice has the following decomposition
\begin{equation*}
K_{ab}=n_an_bK_{\lie}+(g_{ab}-n_{a}n_{b})K_{R}\end{equation*}
where $K_{\lie}=\dot{\lie}/\lie$, $K_{R}=\dot{R}/R$ are two scalar functions of proper radius $l$, and $n^a$ is the outward unit normal to the 2-sphere of fixed radius.

The Hamiltonian constraint \cite[(2.5)]{GO} can be written as
\beq\label{HC}
8\pi\rho=&K_R(K_R+2K_\lie)-(2(RR')'-R'^2-1)/R^2
\eeq
and the radial momentum constraint \cite[(2.6)]{GO} is given by
\beq\label{MC}
4\pi J=&K_R'+R'(K_R-K_\lie)/R
\eeq
where $J=\vec{j_r}\cdot \vec{n}$.

It is clear that the sphere $\Sigma_{l}$ with constant $l$ has mean curvature $k=2R^\prime/R$ in the time slice and the expansion along the proper time direction is $p=2K_R$. The null expansions  with respect to the outgoing rays are
\begin{equation*}
\theta_{+}=2(R'+ RK_R)/R, \ \ \theta_{-}=2(R'-RK_R)/R.
\end{equation*}
We construct an auxiliary function as below
\beq\label{F}
F(l):=-\frac{1}{4}\theta_{+}\theta_{-}R^{3}+R=R(R^{2}K_{R}^{2}-{R'}^{2}+1).
\eeq
By the Hamiltonian constraint (\ref{HC}) and the momentum constraint (\ref{MC}), it turns out that
\beq\begin{aligned}
F^\prime(l)=&R^\prime(R^{2}K_{R}^{2}-{R'}^{2}+1)\\ \nonumber
\ &+R(2RR^\prime K^2_R+2R^2K_RK^\prime_R-2R^\prime R^{\prime\prime})\\ \nonumber
=&4\pi(k\rho+pJ)R^{3}.\end{aligned}\eeq

Consider the outermost future apparent horizon, denoted by $\Sigma_{l_0}$. Let us assume that  $\Sigma_{l_0}$ is
outside the outermost past trapped surface. In other
words, $\theta_+(l_0)=0$ and that both $\theta_+$ and $\theta_{-}$ are
positive outside $\Sigma_{l_0}$.

The dominant energy condition $\rho \geq |J|$ implies that
\beq
\begin{aligned}
\ & 4\pi(k\rho+pJ)\\ \nonumber
=&4\pi(\frac{1}{2}(k+p)(\rho+J)+\frac{1}{2}(k-p)(\rho-J))\\ \nonumber
=&4\pi(\frac{1}{2}\theta_+(\rho+J)+\frac{1}{2}\theta_-(\rho-J))\geq 0 \nonumber
\end{aligned}
\eeq
and therefore the auxiliary function $F(l)$ is monotonically increasing with respect to $l$.
Let $\Sigma_{L_0}$ be the outer boundary, i.e. $L_0 >l_0$. Then we have
 \beq\label{key}F(L_0) \geq F(l_0).
 \eeq
Since $\theta_+(l_0)=0$, the right hand side of (\ref{key}) is the areal radius of the outermost future
apparent horizon $R(l_0)$. The left hand of (\ref{key}) is exactly
$R\left.\Big(1-\big({R'}^{2}-R^2K^2_{R}\big)\Big)\right|_{L_0}$.

Let $\Sigma$ be a spacelike closed 2-surface with spacelike mean curvature vector $\vec{H}$. Recall that the Liu-Yau quasi-local mass for $\Sigma$ is defined by
\begin{equation*}
\ly(\Sigma)=\frac{1}{8\pi}\int_{\Sigma}(k_0-|\vec{H}|)\ud\sigma.
\end{equation*}
It was proved in \cite{LY} that $\ly(\Sigma)$ is nonnegative if the spacetime satisfies the dominant energy condition.
For a round sphere $\Sigma_l$ with areal radius $R(l)$, we have
\begin{equation*}
\begin{aligned}
2\ly(\Sigma_{l})
 =&\frac{2(4\pi R^2)}{8\pi}\frac{2}{R}\left(1-\sqrt{{R'}^{2}-R^2K^2_R}\right)\\
 =&R \left( 2-2\sqrt{{R'}^{2}-R^2K^2_R}\right).
\end{aligned}
\end{equation*}
By means of an elementary inequality $2-2\lambda \geq 1- \lambda^2$, it yields
\begin{equation*}
2\ly(\Sigma_{L_0}) \geq R\left.\Big(1-\big({R'}^{2}-R^2K^2_{R}\big)\Big)\right|_{L_0} \geq R(l_0).\end{equation*}
This is just the localized Penrose inequality (\ref{lpi}) in spherical symmetry for the Liu-Yau mass and the proof is completed.

The underlying ideas constructing the auxiliary function $F(l)$ (\ref{F}) to have the delicate estimate (\ref{key}) are borrowed from and motivated by the work of Malec and \'{O} Murchadha \cite{MO}. In particular, we have managed to established a localized version of the equations (8) and (10) in \cite{MO}. Since we are working within the finite region and we do not need the appropriate decay condition near infinity. But we should emphasize that the current proof here also reveals the spirit of Malec and \'{O} Murchadha and can be viewed as a refined estimate of Malec-\'{O} Murchadha inequality \cite{MO} at a (quasi-)local level.

{\it\noindent Conclusion and Discussion.} We have demonstrated that the localized Penrose inequality is valid for the Liu-Yau quasi-local mass in spherically symmetric geometries. When the time slice is symmetric, the Liu-Yau mass reduces to the Brown-York mass and the future and past apparent horizons coincide. The dominant energy condition reduces to the condition that the finite region $\Omega$ has nonnegative intrinsic scalar curvature.
Then the localized Penrose inequality (\ref{lpi}) becomes
\begin{thm}[Localized Riem. Penrose Inequality]
\beq\label{LRPI}
{\mathbf m}_{\emph{\textrm{BY}}}(\Sigma_{\emph{\textrm{O}}}) \geq \sqrt{|\Sigma_{\emph{\textrm{H}}}|/16\pi}
\eeq
where ${\mathbf m}_{\emph{\textrm{BY}}}(\Sigma_{\emph{\textrm{O}}})$ is the Brown-York mass of the outer boundary $\Sigma_{\emph{\textrm{O}}}$, and $|\Sigma_{\emph{\textrm{H}}}|$ is the area of the outermost apparent horizon.
\end{thm}

Unlike the Riemannian Penrose inequality for asymptotically flat initial data, there is no hope to have a rigidity statement in our theorem. The philosophy of the rigidity is similar to the Liouville type property of the harmonic function on the entire space, i.e. the infinity determines the interior. But this does not hold true for a finite region. Let $\Omega$ be (part of) the spatial Schwarzschild space. Obviously, the areal radius of outermost apparent horizon is $2m$ where $m$ is the Schwarzschild mass. If we take the outer boundary $\Sigma_{\textrm{O}}$ close to the apparent horizon, then the left hand side of (\ref{LRPI}) is approximately $2m$ while the right hand side is exactly $m$. In this case, the localized Riemannian Penrose inequality (\ref{LRPI}) is a strict inequality. On the other hand, our result is also sharp. If we let the outer boundary $\Sigma_{\textrm{O}}$ be a large coordinate sphere, the Brown-York mass approaches to the ADM total mass $m$ \cite{FST}. Then the localized Riemannian Penrose inequality (\ref{LRPI}) becomes an equality and it recovers Malec-\'{O} Murchadha's previous result \cite{MO}.

Recently, Miao also obtained a theorem of quasi-local mass type for a body surrounding horizons \cite[Theorem 1]{Mi} by extending Shi-Tam's positivity argument \cite{ST} and making use of the Riemannian Penrose inequality (\ref{RPI}). He assumed that $\Omega$ has nonnegative scalar curvature, the outer boundary $\Sigma_{\textrm{O}}$ is a round sphere and has positive mean curvature, then
\beq\label{Min}
m(\Sigma_{\textrm{O}})\geq \sqrt{|\Sigma_{\textrm{H}}|/16\pi}\eeq
where the quasi-local type quantity $m(\Sigma_{\textrm{O}})$ is defined as
\begin{equation*}
m(\Sigma_{\textrm{O}})=\sqrt{\frac{|\Sigma_{\textrm{O}}|}{16\pi}}\Big[1-\frac{1}{16\pi|\Sigma_{\textrm{O}}|} \big(\int_{\Sigma_{\textrm{O}}} H \ud \sigma \big)^2  \Big].
\end{equation*}
There is also a comparison theorem \cite[Theorem 3]{Mi} showing that the quantity $m(\Sigma_{\textrm{O}})$ lies between the Hawking mass and the Brown-York mass, i.e. $\by(\Sigma_{\textrm{O}})  \geq m(\Sigma_{\textrm{O}})\geq \hawking(\Sigma_{\textrm{O}})$. In time symmetric case, our localized Riemannian Penrose inequality (\ref{LRPI}) can be viewed as a corollary of Miao's inequality. However, the non-time symmetric version of Miao's inequality (\ref{Min}) and its relation to our result (\ref{lpi}) is still unclear. Moreover, physical properties and significance of  $m(\Sigma_{\textrm{O}})$ certainly repay further study.

A number of interesting questions naturally arises. How to propose an appropriate localized formulation of the Penrose inequality for other quasi-local mass in a spacetime? Even in spherical symmetry? When the cosmological constant appears, what the Penrose inequality with the (anti-) de Sitter reference metric looks like? The spatial geometry in the anti-de Sitter spacetime is modeled by the hyperbolic metric. Various theorems for the positivity of the total mass have been successfully proved \cite{Wa,HPMT}. An associated Riemannian Penrose inequality was conjectured in \cite{Wa} where the apparent horizon was referred as to the constant mean curvature surface. With the hyperbolic reference metric, to define a quasi-local mass becomes much more complicated. One of the so-far-available candidates is the modification of the Liu-Yau mass \cite{WY07} where the closed 2-surface $\Sigma$ is embedded in $\mathbb{H}^3$ with mean curvature $k_0$. Under the dominant energy condition $R\geq -6$, there is a timelike vector $\W^\mu$ such that the mass vector $m_{(\mu)}=\int_\Sigma (k_0-|\vec{H}|)\W^\mu /8\pi$ is causal future. The difficult is that $\W^\mu$ is achieved by solving a backward parabolic equation and hence is not easy to written down explicitly. A similar inequality as shown in this Letter might still be possible for such a mass. The mirrored story for the positive cosmological constant becomes very different. Positive mass theorems hold true only for certain very restrictive initial data sets \cite{LXZ}. It is not easy to extend the Penrose inequality and the techniques here to the asymptotically de Sitter spacetime. Can the physical idea of the formation of black hole and gravitational collapse in the asymptotically de Sitter spacetime be formulated in a mathematically correct way? But clearly this is beyond the scope of this Letter.

Penrose inequality is an important, interesting and challenging problem in physics. No doubt our current knowledge be so far away from the full understanding of this problem. Surely its solution must be very difficult and we require and expect more new both framework and techniques in mathematics as well as physics in the future. Each significant step toward solving this problem is worthwhile and appreciated.

\begin{acknowledgments}
This work is partially supported by the National Science Foundation of China (grants 11171328, 11121101) and the Innovation Program of Shanghai Municipal Education Commission Grant 11ZZ01.
\end{acknowledgments}

\end{document}